# Through Einstein's Eyes

Antony C. Searle, Craig M. Savage, Paul A. Altin,
Francis H. Bennet and Michael R. Hush
*Department of Physics, The Australian National University,
ACT 0200, Australia
craig.savage@anu.edu.au*

## Seeing is believing

Take a ride through the solar system at close to the speed of light. What do you see?

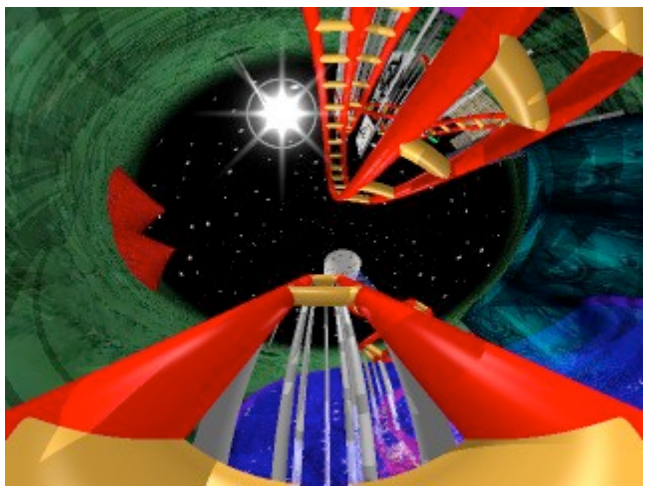

Traditional special relativity teaching sometimes doesn't equip us to answer that question. Length contraction and time dilation are what we measure with a grid of clocks and rulers, not what we would see with our own eyes. The finite speed of light means that we see objects not where they 'are', but where they 'were', adding another layer of distortions on top of the Lorentz transformation. Add in the Doppler shift of objects' colours, and the optical effects of relativity are striking and often beautiful.

We have developed a relativistically-accurate computer graphics code [1, 2] and have used it to produce photo-realistic images and videos of scenes where special relativistic effects dominate, either in astrophysical contexts or in imaginary worlds where the speed of light is only a few metres per second. The videos have been integrated into our undergraduate teaching programme for several years. Recently we took the next step, encouraging undergraduate students to use the code to explore relativity, develop their own videos, and eventually package them together into *Through Einstein's Eyes*, a multimedia CD.

## Highlights

The speed of light – a billion kilometres per hour – means that relativistic visual phenomena in the real world occur on astronomical length scales. One of the scenarios we developed was a grand tour of the solar system, at near-light speeds.

Even though planets are length-contracted as we fly by, they don't look like pancakes – in fact, a result due to Roger Penrose [3] is that spheres always present a circular outline to every observer, regardless of their motion. Even so, every observer will see that circular outline with a different size and in a different part of the sky. Surface details of the planet – like Jupiter's bands or Earth's continents – are bent out of shape by the combination of length contraction and the different times light had to leave different parts of the planet to arrive simultaneously at our eyes. On top of the geometrical distortions, planets are *blueshifted* as we approach them: for example turning Jupiter's clouds from red to blue.

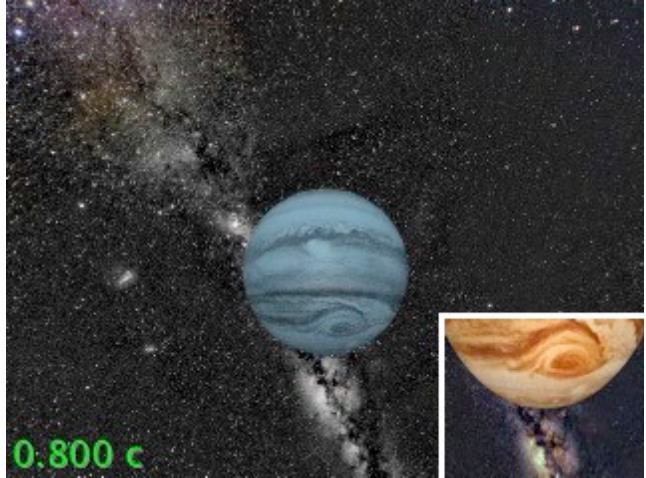

To more clearly see the geometrical distortions produced by relativity, we view objects that aren't spheres, such as Saturn's rings. As we rocket over them, they appear to bend back on themselves!

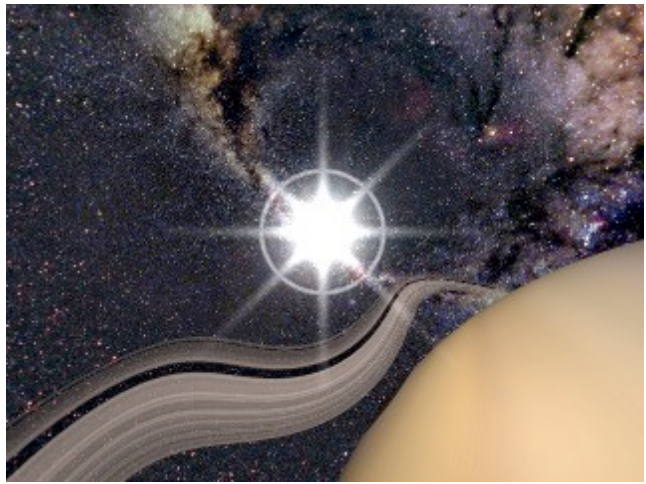

In the tradition of *gedanken* experiments, we imagine a world in which the speed of light is only a few metres per second. Then everyday experiences, like riding a tram or a rollercoaster, take us into a regime of extreme relativistic effects.

Some effects – like curved shadows – can be understood without relativity, as consequences solely of the finite speed of light. In one scene, a tram passing a streetlight casts a curved shadow. The further away from the streetlight the shadow is, the further back in the past the light was blocked by the tram, and the further back along the track the tram was when it blocked the light!

If we ride a rollercoaster at near-light-speed, all kinds of effects come into play. The whole sky shrinks to a circle in front of us, ringed by the horizon. Objects we pass sway and twist. The Doppler effect makes any strong colour (a sharp spectral feature) into a rainbow as we zip by.

Besides the visual experience of relativity, the CD also offers more traditional material such as explanations of time dilation and length contraction using light clocks. For those who are inspired to learn more, the CD contains translations of Einstein's two 1905 relativity papers, and a bibliography and webography. More about the CD, and about the companion DVD, with high quality video for classroom use, is on the Through Einstein's Eyes web site [4].

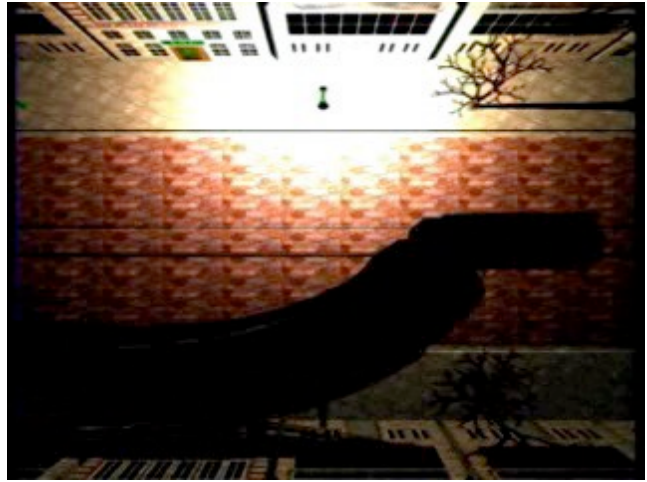

## The Making of Through Einstein's Eyes

In 1997 Antony Searle, then a second year student, created Backlight from scratch in a four day marathon over the Easter long weekend. On Thursday we had just vague ideas; by Tuesday we had a relativistic raytracer. The first video, *Visualising Special Relativity*, featured acceleration down a desert road, a flight through a hollow cube, and Earth orbits. The first two are on the CD. In 1998 Antony rewrote Backlight to allow for objects with different velocities in the same scene. This resulted in the *Seeing Relativity* video, which was explicitly pedagogic in its design, and featured Einstein's tram. Current development of *Backlight* is towards allowing accelerating objects (besides the camera) and improving the illumination model.

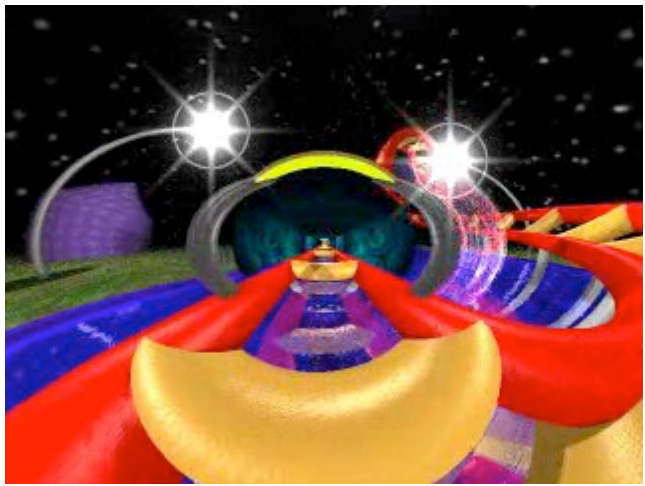

The ANU PhB degree requires students to undertake advanced study courses. Consequently, in 2004 four PhB students, Paul Altin, Michael Hush, Katie Dodds-Eden, and Ben Weise used *Backlight* to create relativistic videos. Supported by an AIP World Year of Physics 2005 grant, Paul and Michael, joined by Francis Bennet, refined the videos and packaged them into the *Through Einstein's Eyes* product. Their reflections provide interesting perspectives on the project…

*Paul*. As our work progressed, we began to understand the need to convey these new and exciting ideas to an audience, to share the things we had learned. It was at this point that *Through Einstein's Eyes* was truly born.

We were exposed to the challenges and opportunities, the successes and setbacks, of research work that most students would not meet until later. As an aspiring physicist I found this to be an excellent opportunity to gain experience in the researching and communication of ideas. The importance of this last component should not be underestimated, and it is often neglected in the training of science students. I would encourage any effort to make such projects a more normal part of undergraduate teaching.

*Michael*. This taught me what a normal course just couldn't. Besides relativistic optics, I was exposed to a type of research environment, which at first was very daunting. In all my other courses each project is designed to teach a specific topic. However, in *Through Einstein's Eyes* we had a very different goal: to produce an original creative product. There was no exact answer; our only guide was to teach relativity to a very wide audience.

I discovered that creative work takes a lot of time. When working on the rollercoaster video it soon became apparent that there was always a way to improve it. The time and effort I put in seemed to increase exponentially as I added the 'finishing touches', which became very frustrating. However, in the end I have my name on a product that I can be proud of, and which is there for the world to see.

*Francis*. I found working with the group encouraged and inspired me to continue to seek out new areas that capture the imagination - and not only in physics. I got some great experience of what it is like to work on a group project, with a subject I could get excited about, and want to teach to others.

We have made a package that shows people what relativity is, in a way that I don't think has been done before. I hope that it will inspire a new way of thinking about physics in general. The idea that physical phenomena can be visualised will surely help to educate and excite a new generation.

## Acknowledgements

This work was supported by the Australian Partnership for Advanced Computing and the ANU Supercomputing Facility, and undertaken on the APAC National Facility. The Australian Institute of Physics supported it as part of its World Year of Physics 2005 activities.